Origins of limited non-basal plasticity in the $\mu$-phase at room temperature

**Authors**
W. Luo[a], C. Gasper[a], S. Zhang[b], P. L. Sun[a], N. Ulumuddin[a], A. Petrova[c], Y. Lysogorskiy[c], R. Drautz[c], Z. Xie[a,*], S. Korte-Kerzel[a,*]

**Affiliation address**
[a] Institute of Physical Metallurgy and Materials Physics, RWTH Aachen University, Kopernikusstraße 14, 52074 Aachen, Germany
[b] Max-Planck-Institut für Eisenforschung GmbH, Max-Planck-Straße 1, D-40237 Düsseldorf, Germany
[c] ICAMS, Ruhr-Universität Bochum, 44801 Bochum, Germany

**\*Corresponding author**
xie@imm.rwth-aachen.de; korte-kerzel@imm.rwth-aachen.de

## Abstract

We unveil a new non-basal slip mechanism in the $\mu$-phase at room temperature using nanomechanical testing, transmission electron microscopy and atomistic simulation. The $(1\bar{1}05)$ planar faults with a displacement vector of $0.07[\bar{5}502]$ can be formed by dislocation glide. They do not disrupt the Frank-Kasper packing and therefore enable the accommodation of plastic strain at low temperatures without requiring atomic diffusion. The intersections between the $(1\bar{1}05)$ planar faults and basal slip result in stress concentration and crack nucleation during loading.

# 1. Introduction

Transition metal topologically close-packed (TCP) phases are potential high-temperature materials owing to their high melting points, good creep resistance, and remarkable strength up to high temperatures [1,2]. However, the notorious brittleness at low temperatures and the limited knowledge of deformation mechanisms, arising from the complex TCP structures, restrict their applications as structural materials [3]. By considering the Nb-Co $\mu$-phase as an archetypal material with a stability reaching from just above the $Nb_6Co_7$ compostion to $Nb_7Co_6$, our recent work [4,5] demonstrates that the plasticity of the TCP phases on the close-packed plane can be tailored by controlling the interplanar spacing and the local shear modulus of the fundamental crystal building blocks. However, the activation of plasticity only on the close-packed (basal) plane is generally insufficient to accommodate arbitrary plastic strain in the complex TCP phases. To improve the ductility, plastic deformation on other planes is necessary. The presence of exclusively tetrahedral interstices necessitates the existence of only four types of Frank-Kasper polyhedra with coordination numbers of 12, 14, 15 and 16 (see Fig S1 in the Supplementary Material) in the TCP structures [3,6]. While small atoms tend to occupy the centers of icosahedral coordination polyhedra (Z12), large atoms prefer to occupy the centers of other coordination polyhedra (Z14, Z15 or Z16), see Fig 1 (a). In case of the $\mu$-phase an exception occurs where the composition is changed in terms of increasing the ratio of the large to small atoms. In this case, when going from $A_6B_7$ to $A_7B_6$ (with A the large and B the small atomic species) the adiditonal A atoms displace the small B atoms inside the characteristic triple layer of the $\mu$- and also Laves phases [4,7].

Different types of faults have been reported in the $\mu$-phases and crystallographically related Laves phases depending on (deformation) temperature and composition. ($\bar{1}100$) prismatic and ($\bar{1}101$) pyramidal planar faults were observed to form in the hexagonal C14 Laves phase after high temperature compression and room temperature impact deformation [8,9]. The coordination polyhedra in these non-basal planar faults remain Frank-Kasper polyhedra although their site occupancies and chemical distributions can differ due to variations in the prevalence of thermally activated processes and chemical compositions [8,10]. Pyramidal planar faults on ($\bar{1}101$) and ($1\bar{1}02$) planes were found in $\mu$-phase precipitates as pre-existing defects [11] and after a creep test at elevated temperature [12]. The formation of those pyramidal faults, which strictly follow TCP packing, was associated with local atomic rearrangements and long-range diffusion. Deformation-induced ($1\bar{1}00$) prismatic planar faults created under quasi-static deformation conditions at room temperature were reported in an off-stoichiometric Mo-Fe $\mu$-phase, where grown-in chemical basal planar defects were widespread [13,14]. While basal slip is frequently seen in $\mu$-phases, non-basal slip is barely reported at low and medium temperatures [5,13,14]. How such plasticity of the µ-phase occurs on non-basal planes at low temperatures, where diffusion is largely suppressed, remains unknown.

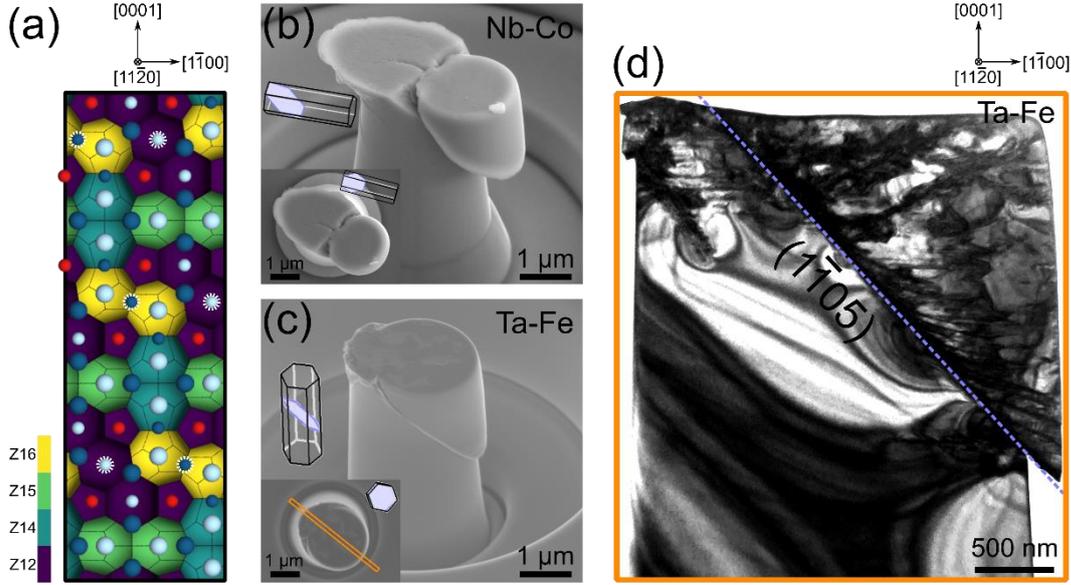

*Fig 1: (a) Atomic structure and coordination environment of a µ-$A_6B_7$ phase. The large and small spheres indicate A and B-type atoms, respectively, and different colors indicate the relative positions of atomic layers along the [11$\bar{2}$0] direction (cf. Fig S2). White dashed circles indicate substitutional sites in A-rich off-stoichiometry [4]. The arrangements of the Frank-Kasper polyhedra were characterized using the 3D Voronoi tessellation. Post-mortem SEM images of (b) a Nb-Co µ-phase micropillar in 88° [0001] orientation (basal plane inclination to the compression axis) and (c) a Ta-Fe µ-phase micropillar in 1.5° [0001] orientation both showing a large slip step parallel to the (1$\bar{1}$05) plane. Top views of the micropillars are inset. Unit cells of the µ-phase showing the corresponding orientations are given next to the micropillars with the orientation of the (1$\bar{1}$05) plane indicated in purple. (d) TEM bright-field image of the Ta-Fe µ-phase micropillar in (c) taken along the [11$\bar{2}$0] zone axis using g = $\bar{1}$10$\bar{2}$. The TEM lamella was cut parallel to the c-axis as illustrated in the inset of (c).*

In this work, we conducted experiments and simulations on several isostructural µ-phases with readily available single phase bulk specimens and/or interatomic potentials of sufficient quality. Nanoindentation and micropillar compression tests were used to activate and investigate non-basal slip in Nb-Co and Ta-Fe µ-phases at room temperature with the Nb-Co phase corresponding more closely to the $A_6B_7$ stoichiometry and the Ta-Fe phase to the $A_7B_6$ stoichiometry. The structure of the non-basal planar fault induced by indentation was imaged by using annular bright-field (ABF) and high-angle annular dark-field (HAADF)- scanning transmission electron microscopy (STEM). The geometrically and energetically favorable metastable states on non-basal slip systems were investigated in detail for the Nb-Ni µ-phase (both $Nb_6Ni_7$ and $Nb_7Ni_6$) using a machine learning interatomic potential. In order to assess the same states also for the Nb-Co and Ta-Fe systems, where no suitable interatomic potentials are available and ab-initio calculations are computationally expensive, we also calculated geometric $\gamma$-surfaces [15] for these two systems. In addition, the transition mechanisms of the non-basal slip events and associated activation energies were calculated for the $Nb_7Ni_6$ phase using the nudged elastic band (NEB) method.

## 2. Methods

### 2.1. Sample preparation

In the present work, a Nb-Co $\mu$-phase alloy with a target composition of 49.5 at.% Nb was prepared from pure Nb (99.9 wt%) and Co (99.98 wt%) by arc-melting in an argon atmosphere and heat treated at 1150 °C for 500 h under high vacuum. A Ta-Fe $\mu$-phase alloy with a target composition of 54.0 at.% Ta was synthesised using an arc melter with Fe (99.99 wt.%) and Ta (>99.99 wt.%) as input materials without heat treatment. Nanoindentation was carried out in the $Nb_{6.4}Co_{6.6}$ (at.%) alloy at room temperature using a nanoindenter (iNano, Nanomechanics Inc.) equipped with a diamond Berkovich indenter tip (supplied by Synton-MDP AG) up to a maximum load of 45 mN. Cylindrical micropillars of 2 μm diameter were cut from the Nb-Co and Ta-Fe $\mu$-phase alloys by focused ion beam (FIB) (Helios Nanolab 600i, FEI) milling.

### 2.2. Mechanical testing

The *in situ* micropillar compression tests at room temperature were conducted inside a scanning electron microscopy (SEM) (CLARA, Tescan GmbH) using an intrinsically load controlled indenter (InSEM, Nanomechanics Inc.) equipped with a diamond flat punch of 10 μm diameter (Synton-MDP AG) at a strain rate of about $10^{-3}$ $s^{-1}$ during stable loading as well as a displacement-controlled indenter (Hysitron PI 89 SEM PicoIndenter, Bruker) equipped with a diamond flat punch of 5 μm diameter with a velocity of about 1 nm/s. The nanoindentation tests and micropillar compression tests of the Nb-Co $\mu$-phase alloy were performed in orientations where *c*-axis of the unit cell of $\mu$-phase is nearly perpendicular to the loading axis. The micropillar compression tests of the Ta-Fe $\mu$-phase alloy were performed in orientations where *c*-axis of the unit cell of the $\mu$-phase is nearly parallel to the loading axis. As basal slip is prohibited in those orientations due to low Schmid factors, plastic deformation on non-basal planes is expected to occur.

### 2.3. Electron microscopy analysis

Based on post-mortem SEM images and electron backscatter diffraction (EBSD) measurements, the slip traces around the indents and on the surfaces of the micropillars were analyzed to estimate the activated slip systems. Afterwards, a lamella was cut from the deformed Ta-Fe $\mu$-phase micropillar, oriented perpendicular to the slip plane and parallel to the slip direction (as illustrated in the inset of Fig 1 (c)). A TEM (JEM-F200, JEOL) operated at 200 kV was used to examine the defects introduced during the compression test. Furthermore, STEM observations were carried out on a lamella cut from an indented area of the Nb-Co $\mu$-phase alloy to identify the defects introduced during indentation using a probe-corrected Titan Themis microscope operated at 300 kV. The probe semi-angle, and the angular ranges of the ABF and HAADF detectors are 24, 8-16, and 73-200 mrad, respectively.

*2.4. Simulation setup*

Atomistic simulations were performed using LAMMPS [16] to calculate the generalized stacking fault energies of non-basal slip systems and corresponding minimum energy paths (MEPs). We chose to perform these calculations on the Nb-Ni $\mu$-phases, which are isostructural to the experimentally considered Nb-Co and Ta-Fe $\mu$-phases (and those containing many different elements as precipitates in most experimental reports in the literature), utilizing a newly developed interatomic potential.

The interatomic interactions were modelled by an atomic cluster expansion (ACE) potential which provides a formally complete basis for the local atomic environments [17]. The ACE for Nb-Ni was parameterized using the pacemaker package [18,19]. Reference data, energies and forces, were generated using the Vienna ab-initio simulation package (VASP) [20–22] with the Perdew-Burke Ernzerhof (PBE) exchange-correlation functional [23].

Spin-polarization was included for Ni and Nb-Ni structures, magnetism was not considered for pure Nb. The energy cutoff for the plane-wave basis was set to 500 eV, and Gaussian smearing with a width of 0.1 eV was applied. The Brillouin zone was sampled using a dense Γ-centered k-point mesh with a spacing of 0.125 Å$^{-1}$. The reference data included various crystalline prototypes with several volumes, surfaces, special quasi-random structures and multiple TCP phases with different compositions.

This dataset was divided into 28 947 structures with 448 055 atoms for the training and 742 structures with 11 584 atoms for the testing. Configurations close to the convex hull were given higher weights in the loss function, while high-energy configurations contributed less, which resulted in a more accurate description of physically relevant structures. The root mean square errors (RMSE) were 8.23 (8.1) meV/atom for energies and 45.19 (52.06) meV/Å for force components for structures within 1 eV/atom above the convex hull for the training and test sets, respectively. The error metrics for the complete training and test datasets were 53.84 (42.52) meV/atom and 191.03 (204.07) meV/Å, respectively. The final ACE potential consists of 1300 basis functions per element and 6208 total trainable parameters. The potential accurately reproduces the formation enthalpies of Nb-Ni $\mu$-phases with various site occupancies obtained using the density functional theory (DFT) method [7] (see Fig S3). The potential properties are presented in Table S1 in the Supplementary Material.

Previous experiments and thermodynamic calculations have demonstrated that the Nb-rich $\mu$-Nb$_7$Ni$_6$ phase is an equilibrium intermetallic Nb-Ni phase [24–26], therefore, most calculations in this work were performed on this phase. The $\mu$-phase crystal structures were constructed using Atomsk [27] and then fully relaxed. The $\gamma$-surfaces were obtained by incrementally shifting one-half of the crystal along the slip direction across the slip plane with periodic boundary conditions (PBC) and then relaxed in the direction perpendicular to the slip plane. Fixed boundary conditions were applied along the slip plane normal, where the outermost atomic layers with a thickness of 14 Å were fixed. The possible planar fault states identified on the $\gamma$-surfaces were further fully relaxed in all directions except for the fixed outermost atomic layers. The FIRE [28,29] algorithm with a force tolerance of $10^{-8}$ eV/Å was used for relaxation.

The transition mechanisms of the non-basal partial slip events were studied using the NEB method [30,31]. The stress-dependent activation energies of slip events were calculated. The initial configuration of the NEB calculation on the partial slip was pre-strained by a transverse displacement corresponding to the partial slip vector. The spring constants for parallel and perpendicular nudging forces are both 1.0 eV/Å$^2$. QUICKMIN [32] was used as the damped dynamics minimizer to minimize the energies across all replicas with the force tolerance of 0.01 eV/Å. All 96 intermediate replicas were initially equally spaced along the reaction coordinate (RC). For more details of the NEB methods and the calculation of activation volume, see [33,34].

The atomic environments of the $\mu$-phase crystals containing planar faults were characterized using the Voronoi tessellation algorithm implemented in OVITO [35]. The Voronoi index was used as a characteristic signature of an atomic cluster, e.g., an icosahedral cluster (Z12) corresponds to the Voronoi index vector (0,0,0,0,12,0,…). Visulization and displacement analysis were carried out using OVITO.

2.5. *Geometric analysis*

The geometrically favorable metastable states of non-basal slip systems for the Nb$_6$Co$_7$ and Ta$_7$Fe$_6$ $\mu$-phases were identified on the geometric $\gamma$-surface [15], which quantifies the changes in atomic volume during slip. The overlapping atomic volume was defined as the absolute deviation of the total atomic volume from the initial crystal during the rigid body shift of a half of the crystal across the interested slip plane. The atomic volume was calculated using the poly-disperse Voronoi tessellation algorithm. The resulting overlapped atomic volumes were normalized by the area of the plane. The input Nb$_6$Co$_7$ and Ta$_7$Fe$_6$ structures were fully relaxed using DFT as detailed in our previous study [4].

# 3. Results

## 3.1. Experiments

We compressed single crystalline micropillars of Nb-Co and Ta-Fe $\mu$-phases in orientation impeding basal glide, that is with the basal plane normal approximately parallel or perpendicular to the compression axis. After compression, the micropillars of the Nb-Co (Fig 1 (b)) and Ta-Fe (Fig 1 (c)) $\mu$-phases with near-zero Schmid factor on the basal plane both exhibit a large and localized slip step parallel to the $(1\bar{1}05)$ plane (Fig 1 (d)).

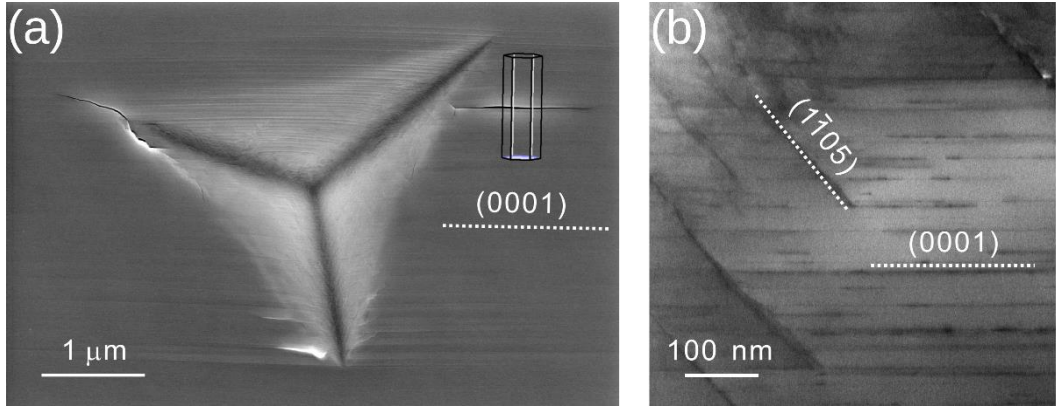

Fig 2: (a) SEM image of an indentation of the Nb-Co $\mu$-phase in 83º [0001] orientation. (b) ABF-STEM image of the indentation showing basal defects and non-basal planar faults parallel to the $(1\bar{1}05)$ plane.

We also performed nanoindentation, which in contrast to microcompression, induces a three-dimensional stress field stressing many different slip systems. A representative SEM image of an indent of the Nb-Co $\mu$-phase, where the loading axis was 83° inclined to the *c*-axis of the unit cell, is shown in Fig 2 (a). The Nb-Co $\mu$-phase exhibits a high density of straight surface traces parallel to the basal plane around the indent as well as a few complex wavy traces and cracks at the corners.

An ABF-STEM image (Fig 2 (b)) taken underneath the indentation shows numerous basal defects and non-basal planar faults parallel to the $(1\bar{1}05)$ plane. The structure of the non-basal planar fault was recorded in a HAADF-STEM image along the $[11\bar{2}0]$ axis (Fig 3 (a)). A basal stacking fault (highlighted by a white dashed box in Fig 3 (a)) exists and intersects with the $(1\bar{1}05)$ non-basal planar fault. The closure failure of the Burgers circuit (indicated in blue in Fig 3 (a)) suggests that the Burgers vector of the partial dislocation bounding the $(1\bar{1}05)$ fault planar is $b = 0.07[\bar{5}502]$. We use a tiling representation to visualize the structural changes across the non-basal planar fault in Fig 3 (b). Along the $(1\bar{1}05)$ plane, tiles of alternating yellow rectangles and parallelograms show the periodicity of $\mu$-phase, whereas tiles of alternating red parallelograms describe the structure of the non-basal planar fault. A schematic illustration of the shear process is shown in Fig 3 (c). The yellow rectangles and parallelograms shear on the (1$\bar{1}$05) plane along the [$\bar{5}$502] direction.

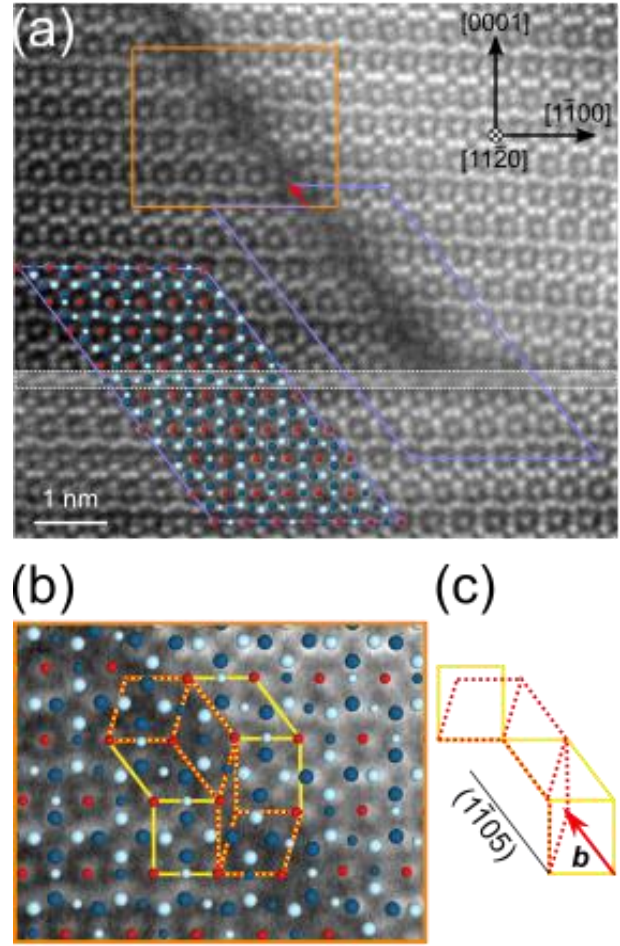

*Fig 3: (a) HAADF-STEM image of a partial dislocation bounding with a (1$\bar{1}$05) planar fault intersects with a basal stacking fault (white dashed box, atomic model overlaid with same color/size scheme as in Fig. 1). The Burgers vector of the (1$\bar{1}$05) partial dislocation is measured from the closure failure of the blue Burgers circuit (note that no closure failure is observed where only the stacking fault is enclosed). (b) An enlarged image of the (1$\bar{1}$05) planar fault in the region marked by an orange square in (a) overlapped with an atomic model.* (c) *Alternating yellow rectangles and parallelograms illustrate the μ-phase crystal lattice and alternating red parallelograms illustrate the structural changes on the (1$\bar{1}$05) plane during deformation. The red arrow indicates the corresponding displacement vector b= 0.07[$\bar{5}$502].*

*3.2. Atomic-scale modelling*

We used energetic and geometric approaches to characterize the formation and atomic structures of the experimentally observed and other potential non-basal planar fault states in the μ-phase. The non-basal slip planes in the μ-phase contain various parallel interlayers (Fig S2) where competing slip mechanisms might occur. The γ-surfaces for the $P_I$ - $P_{VI}$ interlayers of the (1$\bar{1}$05) plane (Fig S2 (c)) in the μ-$Nb_7Ni_6$ phase were calculated (Fig 4 and Fig S4). There are multiple local minima on the γ-surfaces of different $(1\bar{1}05)$ interlayers along the $[\bar{5}502]$ direction with the corresponding displacement vector around $0.07[\bar{5}502]$ (Fig S4), which is consistent with the measured Burgers vector of the $(1\bar{1}05)$ planar fault observed in the HAADF-STEM image (Fig 3). The planar fault state on the $P_{VI}$ interlayer tends to be the most favorable metastable state due to the relatively low energy level and large barriers to the next maxima (Fig S5 (a)). After the partial slip on the $P_{VI}$ interlayer, the coordination polyhedra at the $(1\bar{1}05)$ planar fault remain Frank-Kasper polyhedra (Fig 4 (e)). The stacking fault energy of the $(1\bar{1}05)$ planar fault of the simulated $\mu$-$Nb_7Ni_6$ phase is 197 mJ/m$^2$. A similar potential energy surface of the $(1\bar{1}05)$ slip plane and a $(1\bar{1}05)$ planar fault state with an identical lattice and a stacking fault energy of 220 mJ/m$^2$ were also observed in the $\mu$-$Nb_6Ni_7$ phase as shown in Fig 4 (c,f). Similarly, the geometric $\gamma$-surfaces of the $\mu$-$Ta_7Fe_6$ and $Nb_6Co_7$ phases, particularly the location of the local minima and maxima, qualitatively agree with the corresponding $\gamma$-surfaces of the simulated $\mu$-$Nb_7Ni_6$ and $Nb_6Ni_7$ phases (Fig 4 and Fig S4-8).

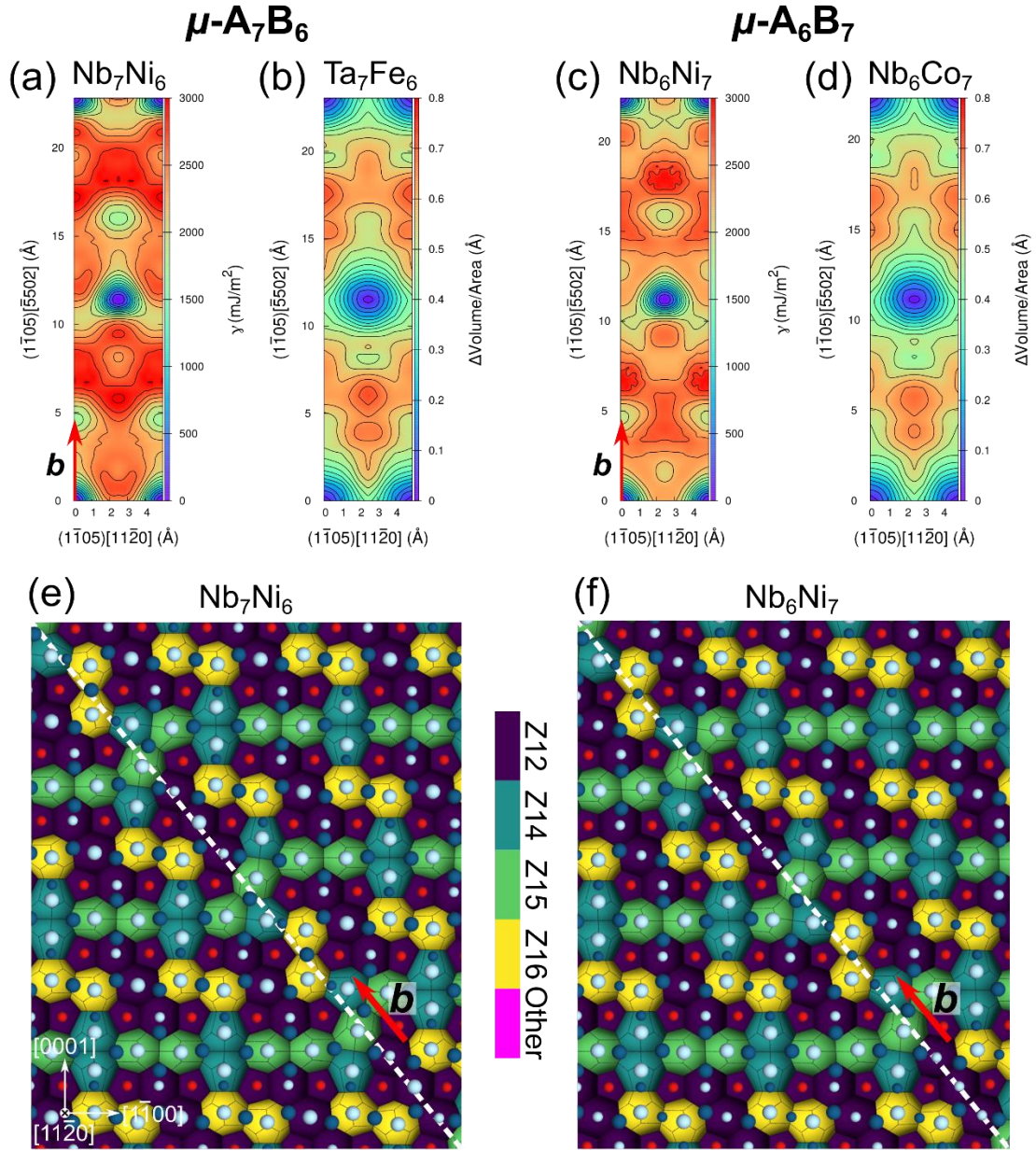

*Fig 4: Glide-induced (1$\bar{1}$05) planar fault state in the μ-phase. γ-surfaces for the $P_{VI}$ interlayer of the (1$\bar{1}$05) plane in the simulated (a) μ-$Nb_7Ni_6$ and (c) μ-$Nb_6Ni_7$ phases. Geometric γ-surfaces for the $P_{VI}$ interlayer of the (1$\bar{1}$05) plane in the (b) μ-$Ta_7Fe_6$ and (d) μ-$Nb_6Co_7$ phases. Atomic structures and coordination environments in the simulated (e) μ-$Nb_7Ni_6$ and (f) μ-$Nb_6Ni_7$ phases after slip on the (1$\bar{1}$05) plane with a displacement vector of 0.07[$\bar{5}$502] and full relaxation. Small and large spheres represent Ni and Nb atoms, respectively. The white dashed lines indicate the slip plane, and the red arrows indicate the displacement vectors. The Frank-Kasper polyhedra Z12, Z14, Z15, and Z16 are colored burgundy, cyan, green, and yellow, respectively. Non-Frank-Kasper polyhedra are colored by magenta.*

Addtionally, we assessed the (geometric) γ-surfaces for different interlayers of the ($\bar{1}$101), (1$\bar{1}$02) and (1$\bar{1}$00) planes in the simulated μ-$Nb_7Ni_6$ phase (Fig S6-8). According to the (geometric) γ-surfaces, the most plausible planar faults on the ($\bar{1}$101), (1$\bar{1}$02) and (1$\bar{1}$00) planes can be formed by ($\bar{1}$101)[1$\bar{1}$02], (1$\bar{1}$02)[$\bar{1}$101] and (1$\bar{1}$00)[0001] slip, respectively. Their respective energy paths are shown in Fig S5 (b-d). The corresponding atomic structures and coordination environments of the possible ($\bar{1}101$), ($1\bar{1}02$) and ($1\bar{1}00$) planar faults formed by partial slip are shown in Fig 5. While non-Frank-Kasper polyhedra are created in the ($\bar{1}101$) (Fig 5 (a)) and ($1\bar{1}00$) (Fig 5 (c)) planar faults, the coordination polyhedra in the ($1\bar{1}02$) planar fault (Fig 5 (b)) remain Frank-Kasper polyhedral. The stacking fault energies of the ($\bar{1}101$), ($1\bar{1}02$) and ($1\bar{1}00$) planar faults in the simulated $\mu$-Nb$_7$Ni$_6$ phase are 639, 525 and 549 mJ/m$^2$, respectively, compared with 197 mJ/m$^2$ on the ($1\bar{1}05$) plane.

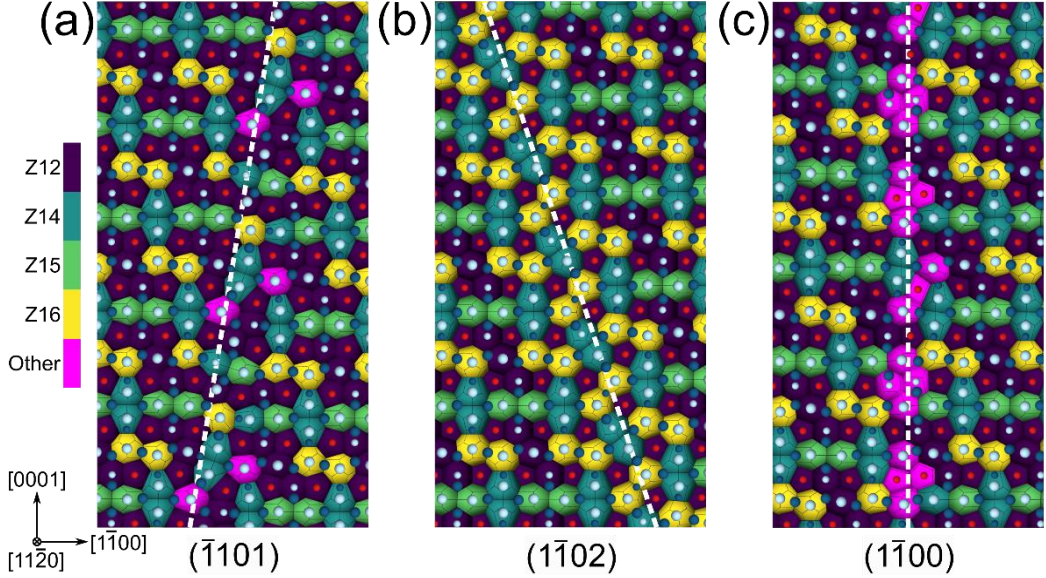

*Fig 5: Atomic structures and coordination environments of the non-basal planar faults in the $\mu$-Nb$_7$Ni$_6$ phase formed by (a) ($\bar{1}101$)[$1\bar{1}02$], (b) ($1\bar{1}02$)[$\bar{1}101$] and (c) ($1\bar{1}00$)[0001] slip. Small and large spheres represent Ni and Nb atoms, respectively, different colors indicate the relative positions of atomic layers along the [$11\bar{2}0$] direction. The white dashed lines indicate the slip plane. The Frank-Kasper polyhedra Z12, Z14, Z15, and Z16 are colored burgundy, cyan, green, and yellow, respectively. Non-Frank-Kasper polyhedra are colored by magenta.*

The minimal energy paths for the formation of the TCP packed ($1\bar{1}02$)[$\bar{1}101$] and ($1\bar{1}05$)[$\bar{5}502$] planar faults in the $\mu$-Nb$_7$Ni$_6$ phase were calculated via the NEB method (see Fig 6). A higher energy barrier of the ($1\bar{1}05$)[$\bar{5}502$] slip (1740 mJ/m$^2$) compared to the ($1\bar{1}02$)[$\bar{1}101$] slip (1436 mJ/m$^2$) was obtained under stress-free conditions. With increasing applied shear stress along the slip direction, the energy barrier of the ($1\bar{1}02$)[$\bar{1}101$] slip decreases less significantly compared to the ($1\bar{1}05$)[$\bar{5}502$] slip (see Fig 6 (b) and energy profiles in Fig S9). A cross-over of the energy barrier between the ($1\bar{1}05$)[$\bar{5}502$] and ($1\bar{1}02$)[$\bar{1}101$] slip occurs around 625 MPa. The calculated activation volume of the ($1\bar{1}05$)[$\bar{5}502$] slip (255 Å$^3$) is almost five times larger than that of the ($1\bar{1}02$)[$\bar{1}101$] slip (56 Å$^3$). The detailed transition mechanisms of slip events and futher interpretation of the results are discussed below.

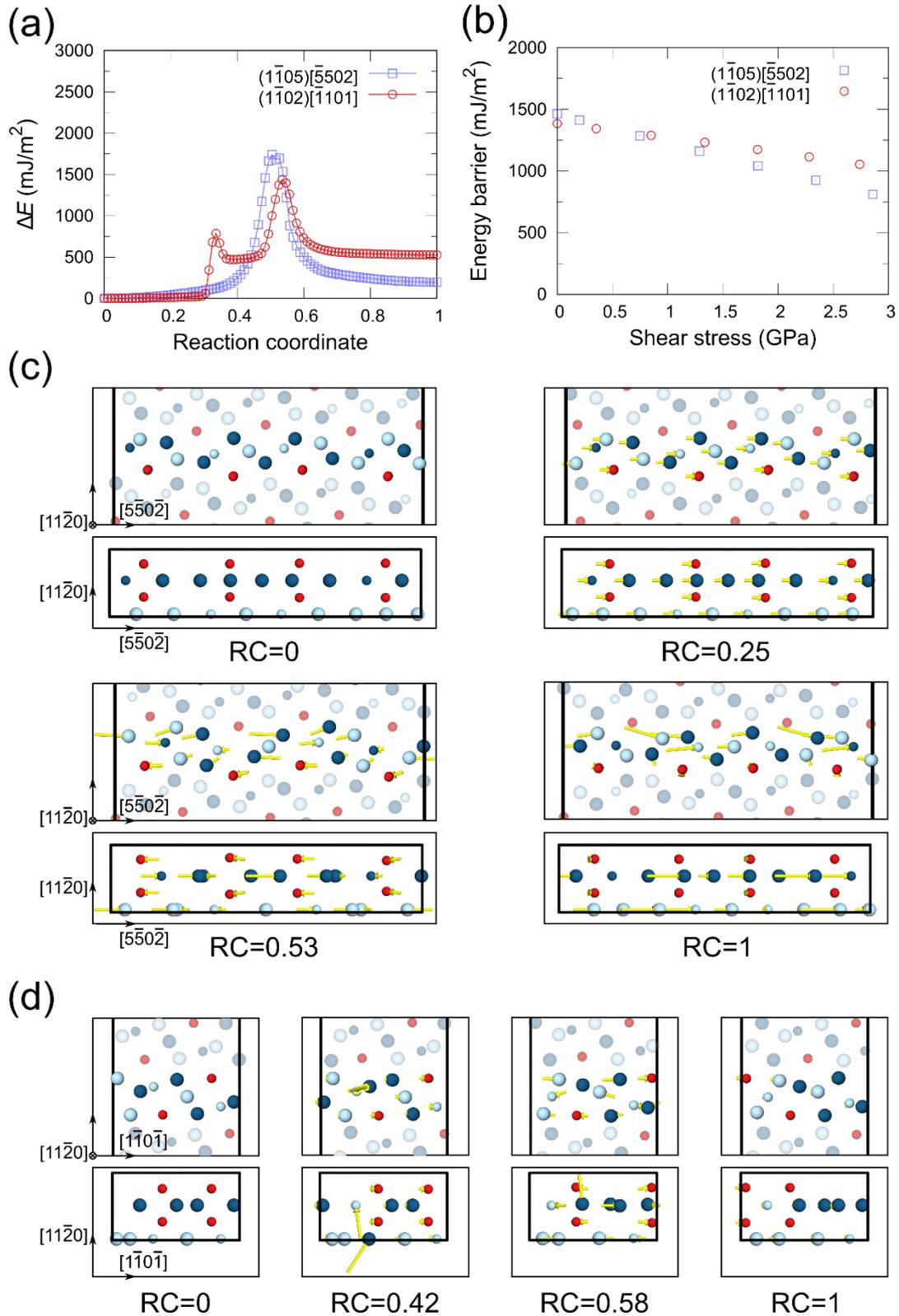

*Fig 6: Minimum energy paths of (1̄105)[5̄502] and (11̄02)[1̄101] slip events in the μ-$Nb_7Ni_6$ phase calculated using nudge-elastic band (NEB). (a) Excess energy vs. reaction coordinate (RC) is calculated using the stress-free setup. (b) Energy barriers of the slip events under varying applied shear stress using the pre-shear setup. Shear stress indicates the stress level applied to the final replica (planar fault state). Transition mechanisms of the (c) (1̄105)[5̄502]*

*and (d) (1$\bar{1}$02)[$\bar{1}$101] slip events. Small and large spheres represent Ni and Nb atoms, respectively, different colors indicate the relative positions of atomic layers along the [11$\bar{2}$0] direction. Only non-transparent atoms in the upper view are displayed in the lower view. The yellow arrows indicate the relative displacement of these non-transparent atoms compared to the previously displayed replica.*

## 4. Discussion

We found experimental and computational evidence for the formation of previously unreported (1$\bar{1}$05) non-basal planar faults in the $\mu$-phases. Here we discuss this finding in terms of two aspects: (1) the associated packing in this and other possible non-basal faults and the changes in the packing necessary during their formation and (2) how this impacts the underlying partial dislocation motion giving a temperature dependence for flow or faults on different non-basal planes.

Pyramidal planar faults on the ($\bar{1}$101) and (1$\bar{1}$02) planes were reported in the Ni-based $\mu$-phase after high-temperature deformation [12]. While the displacement vector of the ($\bar{1}$101) planar fault deviates from the slip plane, the displacement vector ($b$ = 0.09[$\bar{1}$101]) of the (1$\bar{1}$02) planar fault is parallel to the slip plane. Prismatic (1$\bar{1}$00) planar faults were found in the Mo-Fe $\mu$-phase containing widespread chemical basal planar faults [14]. Our simulation results rationalize these experimental observations, indicating that the formation of the stable ($\bar{1}$101) and (1$\bar{1}$00) planar faults involves not only dislocation glide but also atomic transportation with significant thermal assistance to fulfill the TCP packing rule.

The transition mechanisms for the formation of the TCP packed (1$\bar{1}$02)[$\bar{1}$101] and (1$\bar{1}$05)[$\bar{5}$502] planar faults in the $\mu$-Nb$_7$Ni$_6$ phase were illustrated through NEB calculations (see Fig 6). The (1$\bar{1}$02) planar fault can form via a mechanism involving not only shear but also atomic shuffling (see Fig 6 (d)), namely, a short-range atomic transportation in the direction perpendicular to the slip direction, akin to the synchroshear mechanism observed in Laves phases [36]. As thermal assistance is indispensable for the motion of synchro-Shockley dislocations [33,37], it is likely that the formation of the synchroshear-induced (1$\bar{1}$02) planar fault is prohibited at low temperatures even at high stress levels. In contrast, the formation of the TCP packed (1$\bar{1}$05) planar fault via the partial slip does not involve atomic diffusion or significant structural rearrangement (see Fig 6 (c)). Thus, it could occur at low temperatures without substantial thermal assistance, given a sufficiently high applied stress. The different thermally activated natures of the (1$\bar{1}$02)[$\bar{1}$101] and (1$\bar{1}$05)[$\bar{5}$502] slip mechanisms are also illustrated by the distinct activation volumes. At the experimental strength level for $\mu$-phases (a few GPa), the (1$\bar{1}$05)[$\bar{5}$502] slip becomes more energetically favorable than the (1$\bar{1}$02)[$\bar{1}$101] slip. From both geometric and energetic perspectives, the $(1\bar{1}05)$ planar fault is therefore more favourable than the $(\bar{1}101)$, $(1\bar{1}02)$ and $(1\bar{1}00)$ planar faults in the $\mu$-phase at low temperatures.

The activation of the $(1\bar{1}05)$ slip plane was identified experimentally in both isostructural Ta-Fe and Nb-Co $\mu$-phases. The $(1\bar{1}05)$ planar fault structures in the simulated $\mu$-$Nb_7Ni_6$ and $Nb_6Ni_7$ phases align with the HAADF-STEM observations of the Nb-Co $\mu$-phase (refer to Fig 3). In addition, the chemical distribution of atoms along the $(1\bar{1}05)$ planar fault complies with the TCP packing rule, except for one Z12 site (marked by white dashed circles in Fig 1 (a)) occupied by a large Nb atom originally located at a Z12 substitution site in the pristine $Nb_7Ni_6$ $\mu$-phase (Fig S10 (a)). A similar planar fault structure was also obtained in the Ni-rich counterpart ($\mu$-$Nb_6Ni_7$) at the same interlayer and slip vector, where the atomic arrangement strictly follows the TCP packing rule (see Fig S11).

Regarding the $(1\bar{1}02)$ planar fault in the simulated $\mu$-$Nb_7Ni_6$ phase, its atomic structure resembles that observed in a Ni-based $\mu$-phase with 7 alloying elements after creep testing at 1140 °C [12]. Within the structural unit of the $(1\bar{1}02)$ planar fault, a pair of anti-site atoms, which deviates from the TCP packing rule (Fig S10 (b)), suggests that local atomic rearrangements at elevated temperatures, apart from the synchroshear mechanisms, are required to optimize the $(1\bar{1}02)$ fault structure. Given the good correlation among these isostructural $\mu$-phases with different chemical compositions in both experimental and theoretical studies, the geometry factor could serve as a universal indicator of the plasticity in TCP crystals.

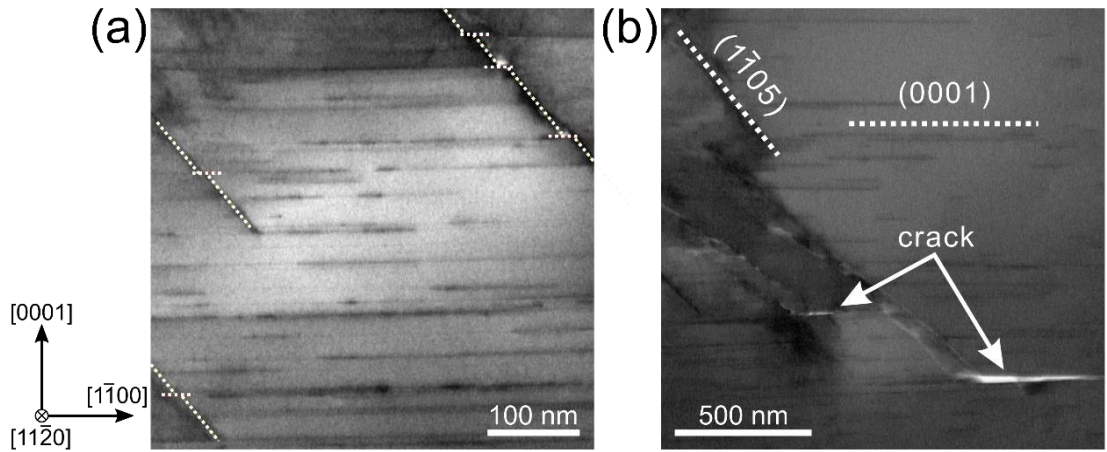

Fig 7: ABF-STEM image of an indentation of the Nb-Co $\mu$-phase in 83° [0001] orientation showing (a) zig-zag $(1\bar{1}05)$ planar faults intersecting with basal defects and (b) basal and non-basal $(1\bar{1}05)$ defects as well as cracks along the basal plane.

Although the $(1\bar{1}05)$ planar fault can be formed by dislocation glide, it remains difficult to operate at low and medium temperatures, requiring high applied stress due to the high energy barrier. In contrast, basal slip in the $\mu$-phase is greatly facilitated by the increased interplanar spacing and reduced stiffness of the slip plane [4], and is therefore more easily and frequently activated. In fact, numerous extensive basal planar defects, presumably growth defects, were observed in our TEM samples. Basal planar faults truncate the pyramidal slip systems and appear to hinder the propagation of the pyramidal dislocations, as the variation of coordination environments caused by the basal planar fault does not allow for the TCP packing of the non-basal planar fault without significant structural rearrangement by atomic diffusion. Similarly, the widespread grown-in chemical basal planar defects in the Mo-Fe $\mu$-phase may suppress the pyramidal dislocation motion [13], thus prismatic plasticity was activated to accommodate $c$-axis strain alternatively. As shown in Fig 7, the glide of basal dislocations is significantly hindered by the $(1\bar{1}05)$ planar faults. The interception of basal slip may significantly increase the stress concentration at the pinning points, leading to dislocation nucleation or crack nucleation during loading. The pinning points could serve as the nucleation sites for the $(1\bar{1}05)[\bar{5}502]$ partial dislocations, as zig-zag $(1\bar{1}05)$ planar faults intersecting with basal defects were widely observed in STEM, see Fig 7 (a). Cracks initiating from the intersections and propagating along the basal plane are shown in Fig 7 (b). This interplay of basal and non-basal defects could be the reason why basal slip governs the plasticity of the $\mu$-phase at room and medium temperatures, as seen from the outside by analysis of the dominant surface traces [4], whereas non-basal plasticity can only be observed by more in-depth analysis but is clearly activated in experiments such as micropillar compression and indentation. In the first, basal slip can be largely suppressed in selected orientations reducing the resolved shear stress and controlling the driving force for fracture using a displacement-controlled compression device (Fig. 1). Underneath an indentation the stress state is complex and contains a hydrostatic component confining deformation and delaying crack formation.

## 5. Conclusions

In this work, we observed experimentally the formation of the $(1\bar{1}05)$ planar fault in different $\mu$-phases and showed that it is created by partial dislocation glide along the $[\bar{5}502]$ direction. We assessed the non-basal slip paths and associated metastable states in different $\mu$-phases using geometric and energetic approaches, revealing that the unique non-basal slip mechanism at low temperatures is determined by the local coordination environments of the $\mu$-phase. This $(1\bar{1}05)$ slip mechanism allows for purely stress-driven deformation without requiring thermal activation to enable diffusional atomic rearrangements. Additionally, the formation energy for the $(1\bar{1}05)$ planar fault is lower than that of the other glide-induced non-basal planar faults. These findings contribute to our understanding of the plasticity in the $\mu$-phase and shed light on the deformation mechanisms of other complex TCP phases at low temperatures, where similar packing-controlled mechanisms are likely to occur.

## Acknowledgments

The authors thank Hauke Springer (IEHK, RWTH Aachen) for the help in sample preparation, Tobias Sedlatschek (IWM, RWTH Aachen) for the help in micropillar compression, Mattis Seehaus (IMM, RWTH Aachen) for the help in conventional TEM characterization, and Thomas Hammerschmidt (ICAMS, RUB) for the help with DFT reference data generation. This project has received funding from the European Research Council (ERC) under the European Union's Horizon 2020 Research and Innovation Programme (Grant Agreement No. 852096 FunBlocks). Funding by the Deutsche Forschungsgemeinschaft (DFG) in the SFB1394 Structural and chemical atomic complexity – from defect phase diagrams to material properties (Project ID 409476157) is gratefully acknowledged. Simulations were performed with computing resources granted by the RWTH Aachen University under Project No. p0020267. AP, YL, RD gratefully acknowledge the funding of this project by computing time provided by the Paderborn Center for Parallel Computing (PC2) for generating reference DFT data and ACE parameterization.

# Supplementary Material

Origins of limited non-basal plasticity in the $\mu$-phase at room temperature

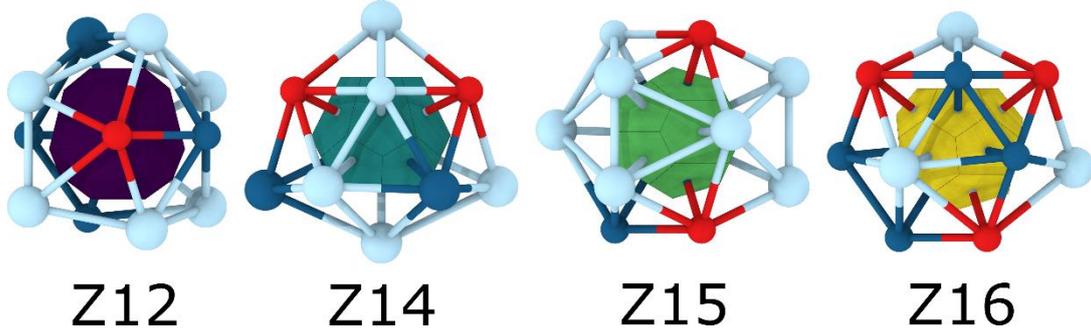

Z12　　　Z14　　　Z15　　　Z16

*Fig S1 Frank-Kasper polyhedra with coordination numbers of 12, 14, 15 and 16 in the TCP structures. The Frank-Kasper polyhedra Z12, Z14, Z15, and Z16 are colored burgundy, cyan, green, and yellow, respectively. The ball-stick models show the atomic configurations of the Frank-Kasper polyhedra.*

*Table S1 Potential properties of stoichiometric $Nb_6Ni_7$ and Nb-rich $Nb_7Ni_6$ $\mu$-phases using the Nb-Ni atomic cluster expansion potential. $a_0$ and $c_0$: lattice parameters; d: interplanar distance; $C_{ij}$: elastic constants; $\gamma_{SF}$: stacking fault energy.*

| Properties | $Nb_6Ni_7$ | $Nb_7Ni_6$ |
|---|---|---|
| $a_0$ (Å) | 4.908 | 4.974 |
| $c_0$ (Å) | 26.068 | 26.616 |
| $d_\text{triple-kagme}$ (Å) | 1.662 | 1.805 |
| $d_\text{triple}$ (Å) | 0.306 | 0.327 |
| $d_\text{kagme-CN14}$ (Å) | 1.112 | 1.030 |
| $d_\text{CN14-CN15}$ (Å) | 1.211 | 1.325 |
| $C_{11}$ (GPa) | 286.9 | 268.1 |
| $C_{12}$ (GPa) | 163.0 | 176.7 |
| $C_{13}$ (GPa) | 162.9 | 166.7 |
| $C_{14}$ (GPa) | 2.0 | -3.2 |
| $C_{33}$ (GPa) | 306.1 | 291.0 |
| $C_{44}$ (GPa) | 55.6 | 50.0 |
| $\gamma_{SF}^{(0001)}$ (mJ/m$^2$) | 41 | 31 |
| $\gamma_{SF}^{(1\bar{1}05)}$ (mJ/m$^2$) | 220 | 197 |

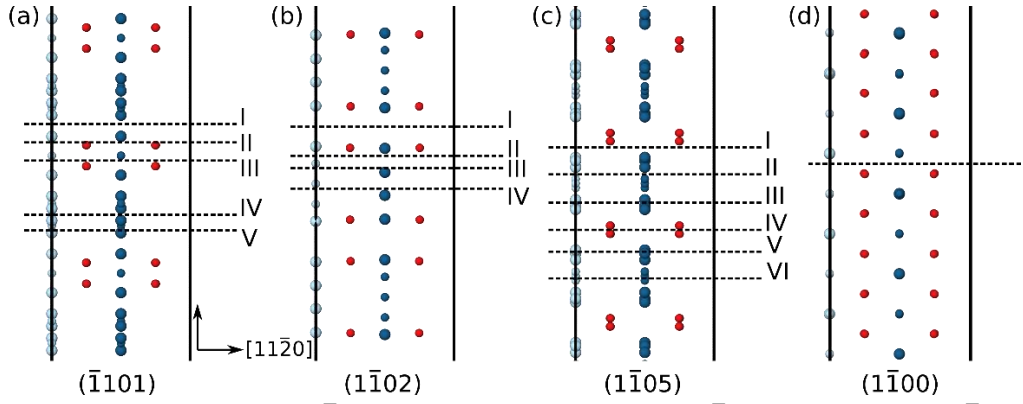

*Fig S2 Atomic structures of (a) $(\bar{1}101)$ plane viewed along the $[1\bar{1}02]$ direction, (b) $(1\bar{1}02)$ plane viewed along the $[\bar{1}101]$ direction, (c) $(1\bar{1}05)$ plane viewed along the $[\bar{5}502]$ direction, and (d) $(1\bar{1}00)$ plane viewed along the $[0001]$ direction. $P_I$ - $P_{VI}$ denotes various interlayers parallel to the slip plane. The large and small spheres represent A and B atoms in $A_6B_7$, respectively, different colors indicate the relative positions of atomic layers along the $[11\bar{2}0]$ direction.*

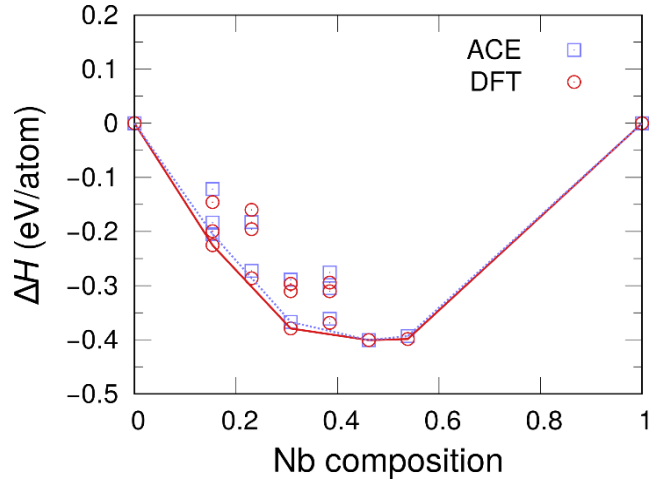

*Fig S3 Formation enthalpies of Nb-Ni μ-phases with different site occupancies calculated using the atomic cluster expansion (ACE) potential and replotted from the density functional theory (DFT) data [M.H.F. Sluiter et al., Physical Review B, 67, 174203 (2003)]. The formation enthalpy of a structure ΔH is defined as its enthalpy H minus the concentration weighted enthalpies of pure Nb and Ni with the μ-phase structure, $\Delta H = H - C_{Nb}H_{Nb} - C_{Ni}H_{Ni}$.*

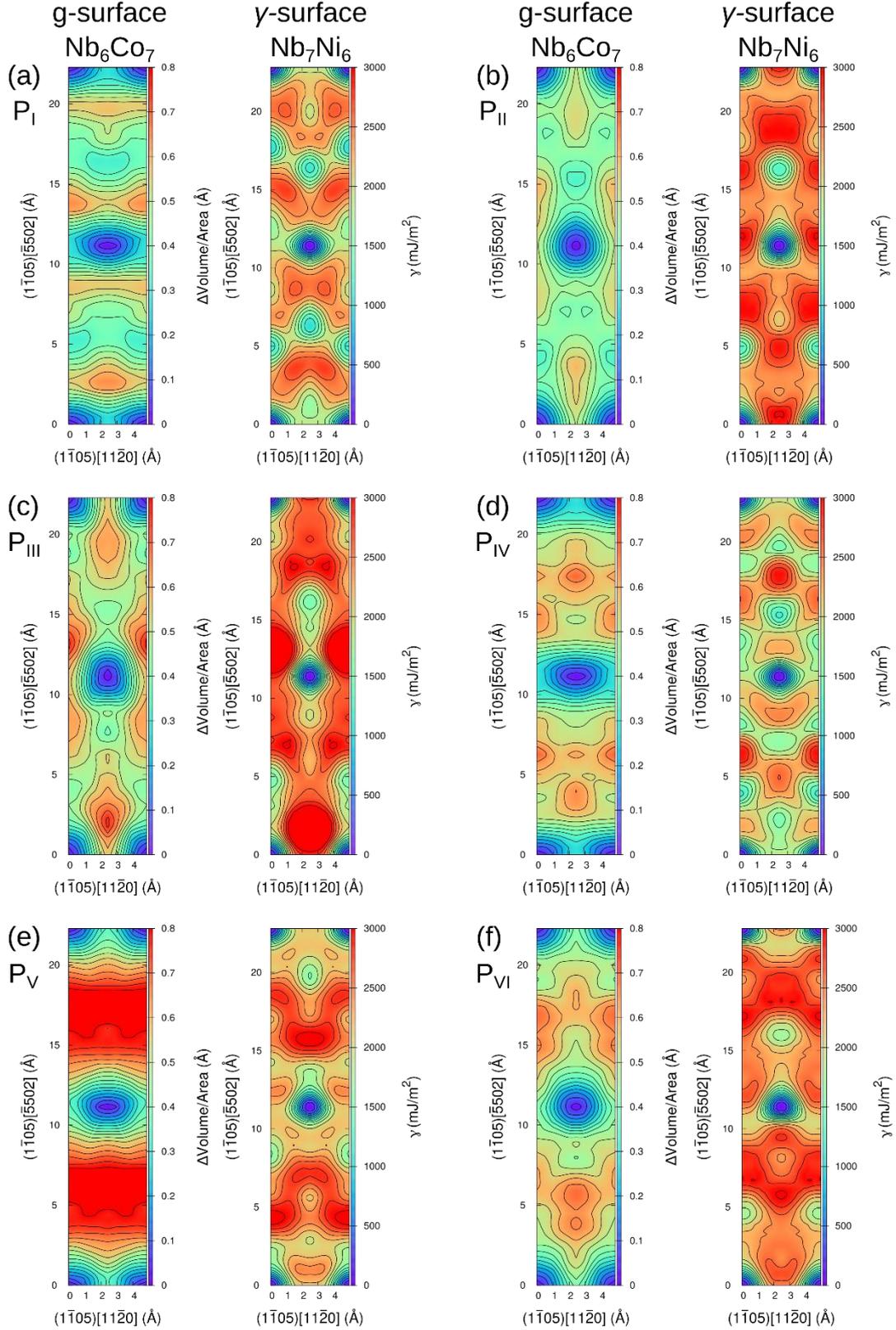

*Fig S4 Geometric γ-surfaces (g-surfaces) of the (1̄105) plane in $Nb_6Co_7$ (left) and γ-surfaces of the (1̄105) plane in $Nb_7Ni_6$ (right) for (a) the $P_I$, (b) $P_{II}$, (c) $P_{III}$, (d) $P_{IV}$, (e) $P_V$, and (f) $P_{VI}$ interlayers.*

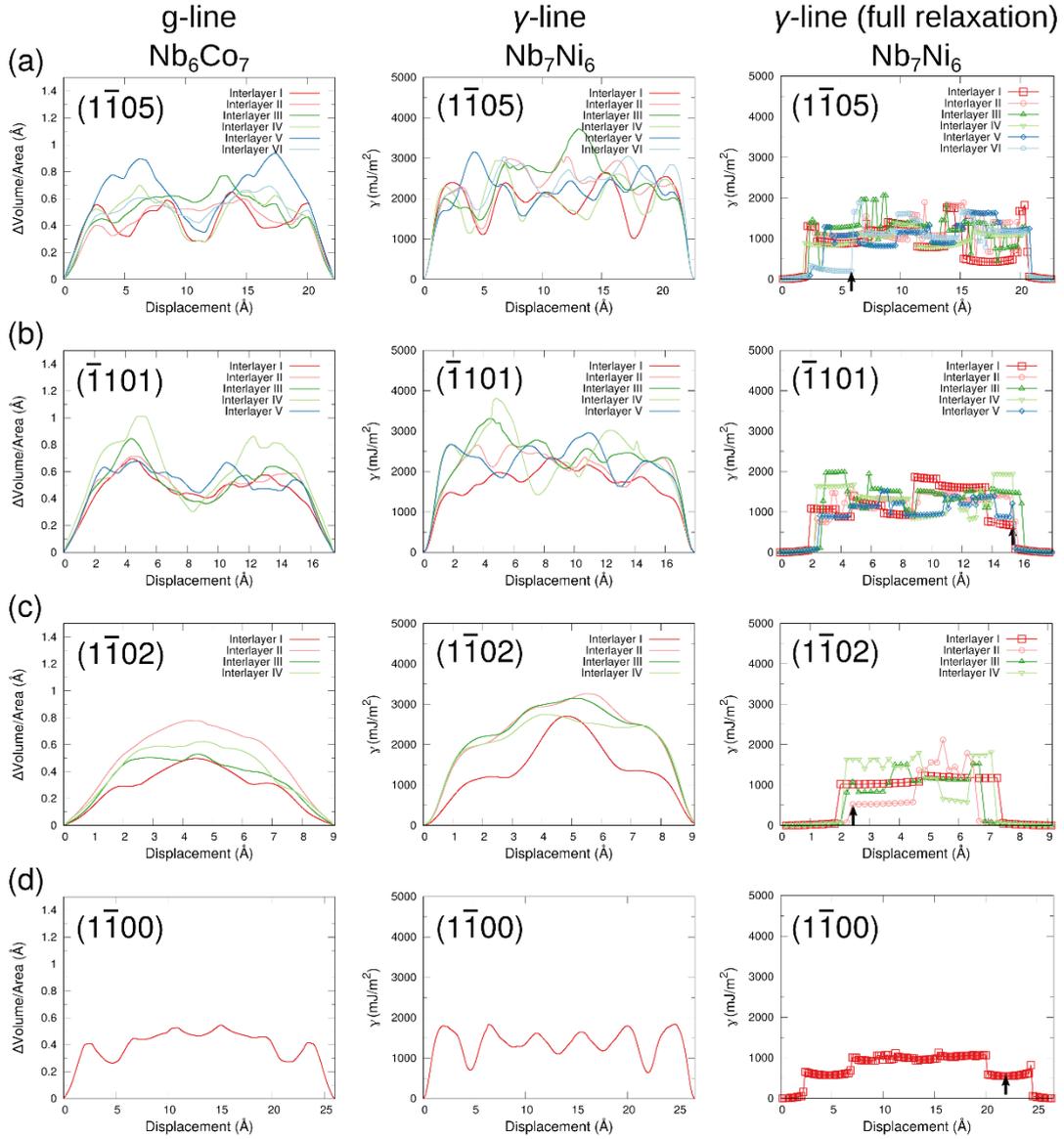

*Fig S5 Geometric γ-lines (g-lines) of the μ-$Nb_6Co_7$ phase and generalized stacking fault energies (γ-lines) and fully relaxed γ-lines of the μ-$Nb_7Ni_6$ phase of (a) $(1\bar{1}05)[\bar{5}502]$, (b) $(\bar{1}101)[1\bar{1}02]$, (c) $(1\bar{1}02)[\bar{1}101]$, and (d) $(1\bar{1}00)[0001]$ slip. The arrows represent identified planar fault states after full relaxation.*

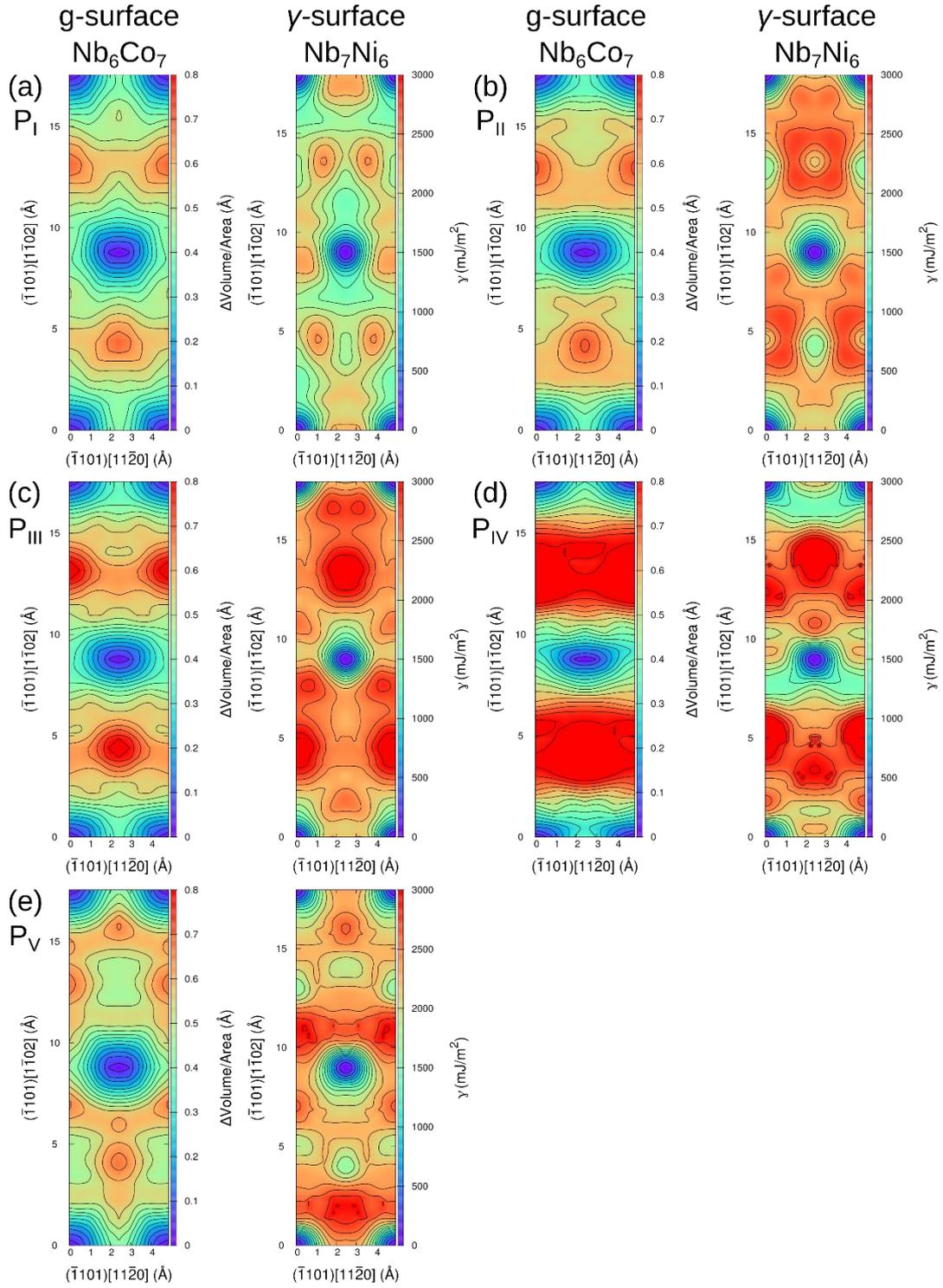

*Fig S6 Geometric γ-surfaces (g-surfaces) of the ($\bar{1}$101) plane in $Nb_6Co_7$ (left) and γ-surfaces of the ($\bar{1}$101) plane in $Nb_7Ni_6$ (right) for (a) the $P_I$, (b) $P_{II}$, (c) $P_{III}$, (d) $P_{IV}$, and (e) $P_V$ interlayers.*

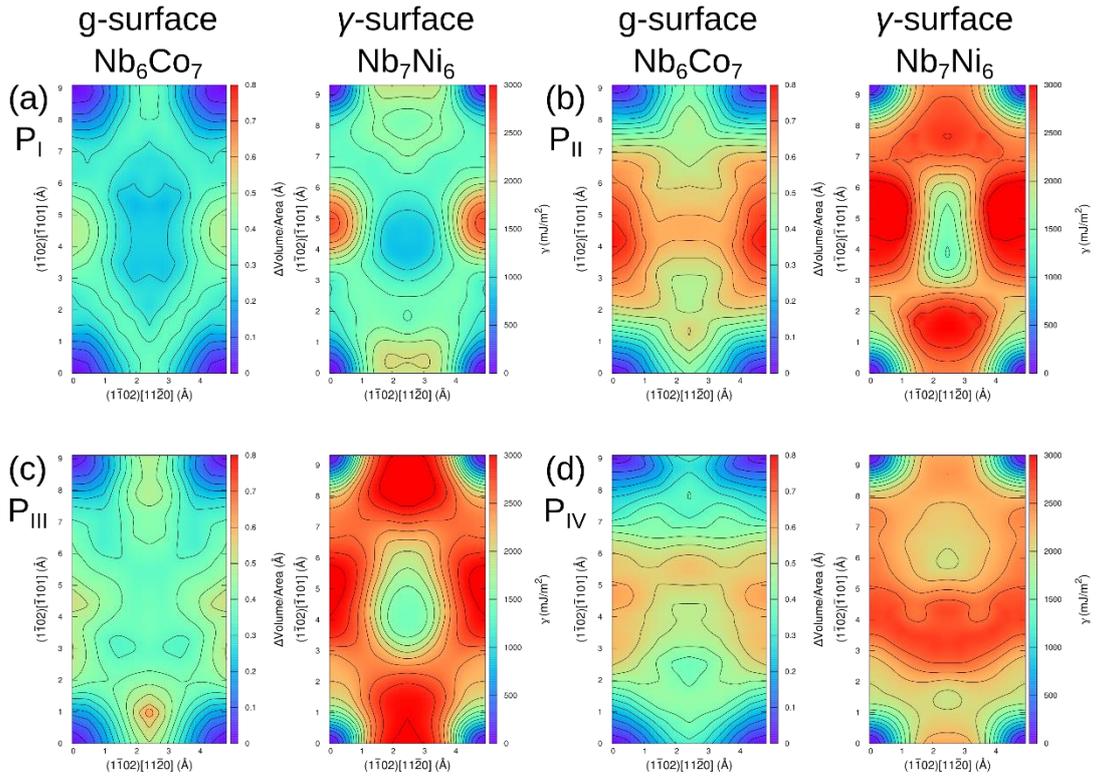

*Fig S7 Geometric γ-surfaces (g-surfaces) of the (1̄102) plane in Nb$_6$Co$_7$ (left) and γ-surfaces of the (1̄102) plane in Nb$_7$Ni$_6$ (right) for (a) the P$_I$, (b) P$_{II}$, (c) P$_{III}$, (d) P$_{IV}$ interlayers.*

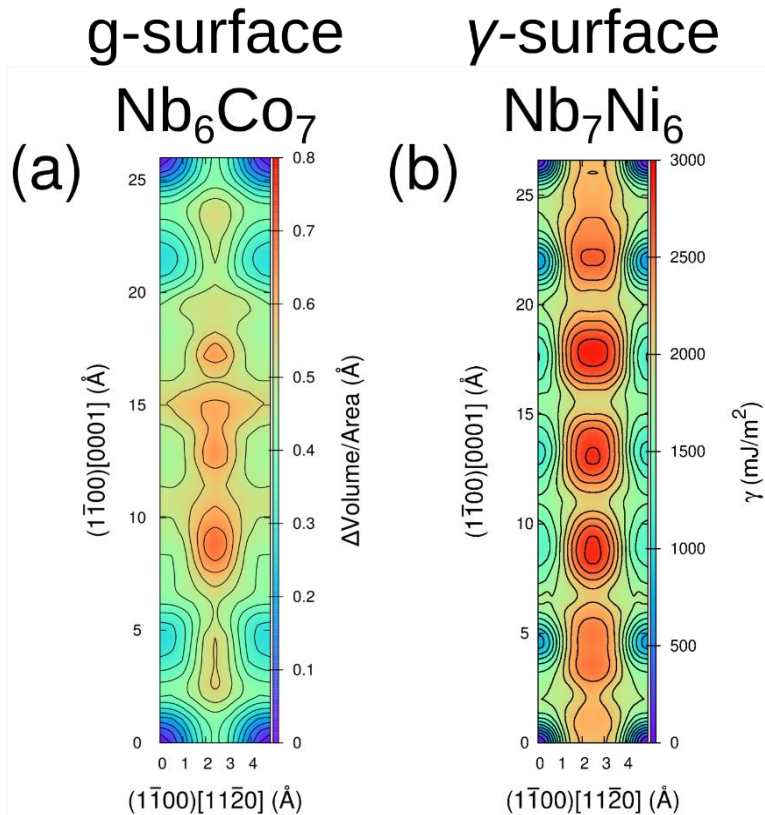

*Fig S8 (a) Geometric γ-surface (g-surface) of the (1̄100) plane in Nb$_6$Co$_7$. (b) γ-surface of the*

($1\bar{1}00$) plane in $Nb_7Ni_6$.

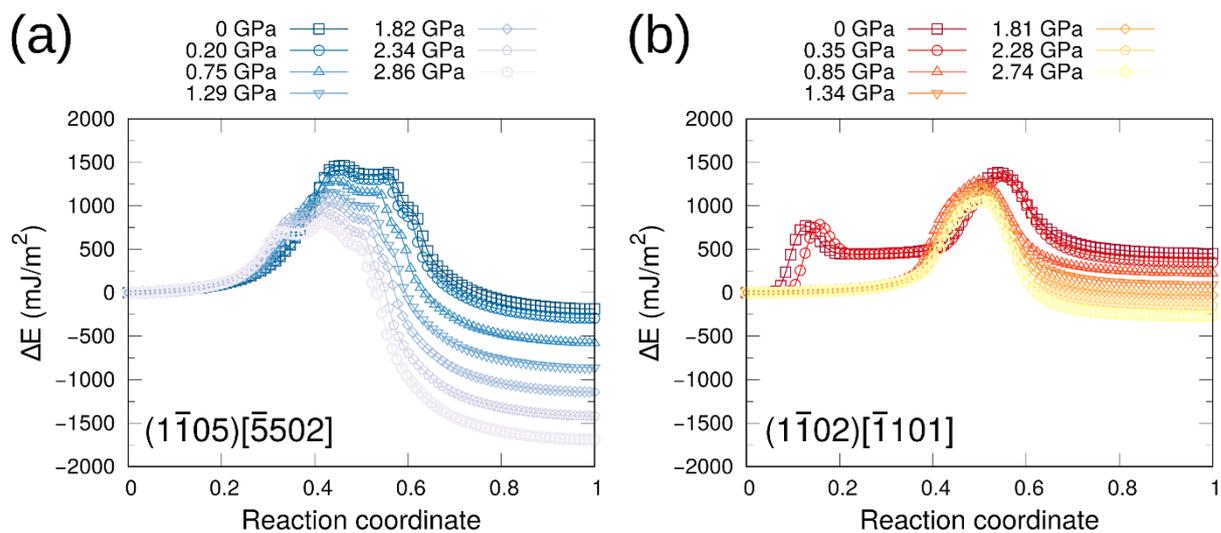

Fig S9 Excess energy vs. reaction coordinate is calculated using the pre-shear nudged elastic band setup for the (a) ($1\bar{1}05$)[$\bar{5}502$] and (b) ($1\bar{1}02$)[$\bar{1}101$] slip.

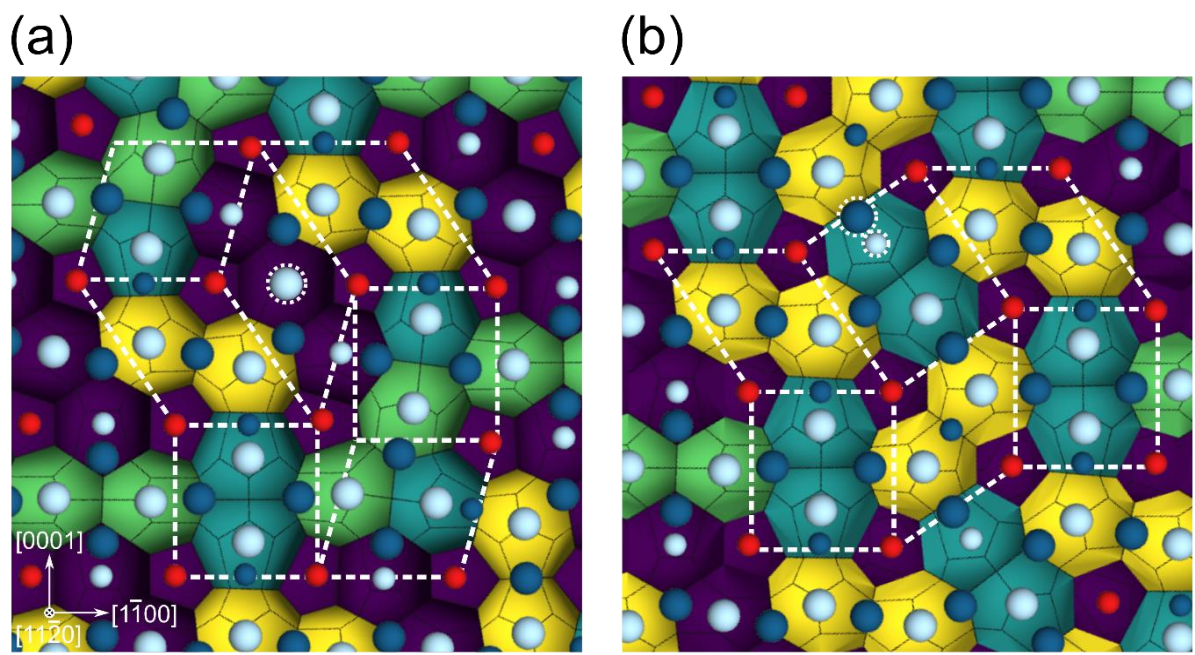

Fig S10 (a) Atomic structure of the glide-induced ($1\bar{1}05$) planar fault in $Nb_7Ni_6$ after full *relaxation. (b) Atomic structure of the synchroshear-induced ($1\bar{1}02$) planar fault in Nb$_7$Ni$_6$ after full relaxation. The small and large spheres represent Ni and Nb atoms, respectively, different colors indicate the relative positions of atomic layers along the [$11\bar{2}0$] direction. White dashed circles indicate anti-sites in planar faults. The Frank-Kasper polyhedra Z12, Z14, Z15, and Z16 are colored burgundy, cyan, green, and yellow, respectively. Non-Frank-Kasper polyhedra are colored by magenta.*

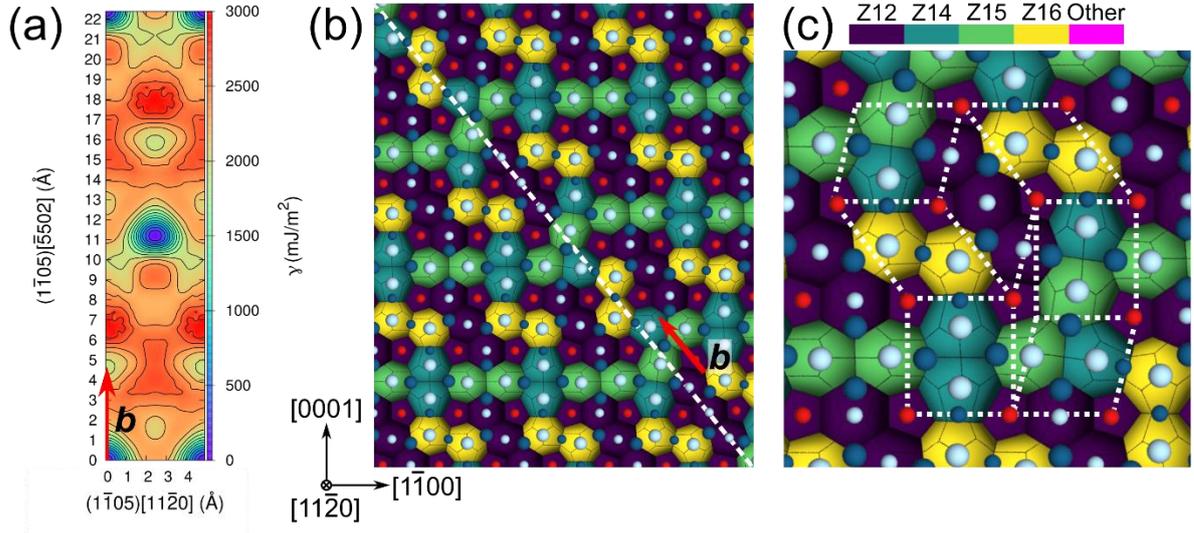

*Fig S11: (a) γ-surface for the P$_{VI}$ interlayer of the ($1\bar{1}05$) plane in the simulated μ-Nb$_6$Ni$_7$ phase. (b) Atomic structures and coordination environments in μ-Nb$_6$Ni$_7$ after slip on the ($1\bar{1}05$) plane with a displacement vector of 0.07[$\bar{5}502$] and full relaxation. The white dashed lines indicate the slip plane, and the red arrows indicate the displacement vectors. (c) Zoom-in view. Small and large spheres represent Ni and Nb atoms, respectively. The Frank-Kasper polyhedra Z12, Z14, Z15, and Z16 are colored burgundy, cyan, green, and yellow, respectively. Non-Frank-Kasper polyhedra are colored by magenta.*